\documentclass[pdflatex,sn-mathphys]{sn-jnl}

\jyear{2021}%

\theoremstyle{thmstyleone}%
\theoremstyle{thmstyletwo}%

\theoremstyle{thmstylethree}%

\usepackage{times}
\usepackage{soul}
\usepackage{float}
\usepackage{setspace}
\onehalfspace 
\renewcommand\rightmark{\textit{Network analysis on political election}}
\renewcommand\leftmark{\textit{Andrea Russo, Vincenzo Miracula, Antonio Picone}}

\begin{document}
\title[]{Network analysis on political election; populist vs social emergent behaviour} 
\author*[1]{\fnm{Andrea} \sur{Russo}}\email{Andrea.russo@phd.unict.it}

\author[1]{\fnm{Vincenzo} \sur{Miracula}}

\author[1]{\fnm{Antonio} \sur{Picone}}

\affil[1]{\orgdiv{Department of Physics and Astronomy "E. Majorana"}, \orgname{University of Catania}} 


\abstract{Social networks play an important role in people's daily socialization, particularly through social media platforms, which have become key channels for communication and information dissemination. The digital ecosystem does not only evolve communication on multi-network (like TV, social media, and online newspapers) but also provides the social researcher with useful data to explain social-complex dynamics. Our work focus on cultural dynamics-reactions that occurred during the 2020 Emilia-Romagna elections'' in Italy, where a stronghold culture felt in danger of losing against the strong populism and Euro-scepticism present in digital ecosystems. We would like to show how the interaction between parts of the society, during cultural and/or political shifting, can lead to or induce emerging behaviour from society, creating groups that react against or improve the status quo. We developed a word-entry network based on three different levels of participation: pro, con, and neutral. We have analyzed the tweets collected (as text) with the word embedding tools, to see, the most used words (which may suggest the main topics) and the most related words among the various groups. We show how a careful analysis of groups through networks, can give important information about the current event.}


\maketitle

\newpage

\section{Introduction} 

Millions of people interact and keep themselves informed by using the Internet. Everyone - somehow - is part of these news systems and is changing the communication tools. Blogs, social networks, and online newspapers are now social relationship tools \cite{Bennato}. 
However, in the 2020s, strong and weak mutations appear, which modify the society and make it react in different ways. Wars, pandemics, and cultural dynamics processes have drastically changed the sociological ecosystem. 

Our work focus on cultural dynamics-reactions that occurred during the 2020 ``Emilia-Romagna elections'' \cite{cairney2012complexity} in Italy, where a stronghold culture \cite{perez2015persistence} felt in danger of losing against a strong populism and Euro-scepticism present in digital ecosystems. 
This election was also important for the national equilibrium of the national government, because, the left party affiliated with the stronghold culture is in the government itself. Losing the election in his most important region can also induce losing the position in the national government (because they are no longer powerful and important) and let the government fall. So, this election was important not only in a regional context but also at national and international levels. 

Inside these national and political dynamics, we would like to show how the interaction between parts of the society, during cultural and/or political shifting, can lead to or induce emerging behaviour from society, creating groups that react against or improve the status quo. The cultural and/or political shifting, is related to the growth of the right-wing populist party, in a left-wing stronghold area, while during the election campaign, the populist party has made a series of populist actions in the physical and digital ecosystem. 

Those actions have obtained powerful mediatic attention, because some of them were also related to the drug cartel of the city (asking people if they were drug dealers), violating people’s private sphere and inducing hate against someone (especially black or migrant), many of this action were also linked to legal and security aspects. 

The research question that we would like to answer is related if it is possible to understand the reason that led to the creation of the emerging group, and if it is possible via computer science to understand the dynamics with the other groups (left and right wings party). We think that the emerging group, called “Le Sardine” (The Sardines), arises due to the fear of losing in that stronghold area, and to react against a political populist party that grows in political support. 

We aim at studying this social phenomenon by building a network to evaluate the interaction between the stronghold culture and the arising populism culture in Emilia-Romagna. 

\section{Methodology}  
It is known that in order to study any kind of phenomenon data are needed, whatever the subject that is being considered is. With the information gathered, we developed a word-entry network based on three different levels of participation: pro, con, and neutral. 

The data was then further processed using \textit{NetworkX} and \textit{Gephi}, the former being one of the most widely used libraries for generating, manipulating, and visualizing networks through \textit{Python}, while the latter is an open source program for analyzing and visualizing social networks.

\subsection{Data Gathering}

 In order to get the data for this analysis we decided to collect a significant amount of textual data\cite{mitchell2018web} over time and from a specific social network: \textit{Twitter}. 

The reasons behind this choice are fairly easy to understand. Twitter is the only social network from which you can easily get data, as it provides a regular \textit{RESTful API}, which is a kind of API (\textit{Application Program Interface}) that is designed to exchange data over the Internet. It requires an authentication method, that was requested through the Twitter developer portal\cite{makice2009twitter}.

In order to consume these APIs we used a library that can be used through Python code: \textit{Tweepy}. It allows developers to interact with the Twitter API in an easier way than making raw HTTP requests, so it is handy to handle big amounts of data.

There is another feature that was useful for our purpose, you can filter your query while using the Twitter API, language, location and hashtags can be chosen. We decided to search all the tweets containing the hashtags \#bolognanonsilega, \#sardine, and \#salvini. 

As already mentioned, additional filters were applied to our search query, not only we selected the above-mentioned keywords but also we specified a geofence, and we collected all the tweets geotagged\cite{davis2011inferring} in a range of 50 km from the city of Bologna, which means that all the tweets in a radius of 50 km from the coordinates of the city center of Bologna were scraped and saved in our database as we wanted to study this phenomenon in a controlled way.

We ended up with a dataset consisting of $\sim$2100 tweets, ranging from 12-11-2019 to 20-01-2020 and they were stored in CSV format (\textit{comma-separated values}). It is a text-based format where each column is separated by either a comma, semicolon, or tab and can be imported into popular spreadsheet editing software like Microsoft Excel. We then had all the data but the informative value was missing for our research yet, as we aimed at understanding whatever each tweet was against the leader of the populists, neutral or pro. We tried to build a machine learning algorithm that is able to distinguish and categorize these 3 categories but it did not give any relevant results.

The main reason for this failure is that a lot of tweets were either ironic or they needed a background context to understand what they were referring to. The only way to accomplish that - despite being more time-consuming - was manually reading each tweet and tagging whenever each one of them was against, neutral, or pro and we used - for the sake of simplicity - respectively 0, 1, or 2.


\subsection{Network \& Community detection} 

Networks are defined as structures \cite{albert2002statistical} that represent the group of objects and/or people, along with their relationship. Social networks play an important role in people's daily socialization \cite{lazer2009computational}, particularly through social media platforms, which have become key channels for communication and information dissemination. It becomes increasingly important to understand how information spreads within these networks. For example, the speed at which information is shared, how users group themselves by similar interests, how they interact with each other, and so on. 

One way to study social networks is through the use of social network analysis (SNA), which is a field of study that involves the use of statistical and mathematical techniques to analyze and understand the relationships and patterns within a network of individuals or organizations. SNA is often used to identify key players and understand the dynamics of social, professional, and communication networks. The aim of SNA is to understand a community by mapping the relationships that connect them as a network and then try to draw out key individuals, and groups within the network and/or associations between the individuals. Twitter is a great platform for observing these behaviours. 

In SNA, a network is typically represented as a graph, with nodes representing the individual actors and edges representing the relationships between them. A variety of different metrics (see Table \ref{tab:1}) can be used to analyze these networks, including measures of centrality (e.g., degree, betweenness), measures of community structure (e.g., modularity), and measures of network evolution \cite{albert2002statistical} (e.g., preferential attachment).

\begin{table} [ht]
    \begin{tabular}{ |l|| l|}
    \hline
    Statistic & Explanation\\
    \hline
    N. of nodes & N. of people in
    the network \\
    N. of links  & N. of relationships
    between people in the network \\
    N. of components & N. of discrete groups
    in the network \\
    Density & The proportion of all links that are actually present \\
    Diameter & Size of the network \\
    Mean avg distance & Avg n. of steps needed to go from one node to any other\\
    Mean degree & Avg n. of links that pass through the nodes \\
    Mean betweenness & Avg n. of unique paths that pass through the nodes\\
    \hline
    \end{tabular}
    \vspace{5pt}
\caption{Statistic and explanation }
\label{tab:1}
\end{table}

In these large social networks, especially linked to important issues such as political elections, people group according to opinions by interacting with some users. We will use Python's implementation for the Louvain community discovery algorithm. The algorithm is described in the work of Blondel et al.\cite{blondel2008fast}. The idea of the algorithm underlies the calculation of modularity, as the metrics used to measure the quality of the division. The modularity measure is a value in the interval [-1, 1] and measures the density of links within communities compared to links between communities.


Controversial topics can be more engaging because they tend to arouse strong emotions and opinions in people. When people encounter information or ideas that challenge their beliefs or values, they may feel motivated to engage with the content and express their opinions. This can lead to increased activity and engagement on social media platforms, as people comment, share, and like content related to the controversial topic.

In their work, Recuero et al.\cite{recuero2019using} conducted an SNA to identify polarized political conversations on Twitter. To understand them, they used several metrics, such as in-degree, out-degree, and modularity. According to the authors, we use modularity as a measure of the strength of the division of a network into distinct groups or communities.

Furthermore, controversial topics can generate a sense of urgency or importance, as people may feel that the topic has immediate relevance or consequences for themselves or others. This, too, can contribute to increased involvement and participation.

It is worth noting that controversial topics can also generate negative or harmful effects, such as misinformation and the spread of fake news. It is important to approach controversial topics carefully and consider the potential consequences of involvement in such content.

\subsection{Word-embedding}
Another tool used in this paper is word embedding. The word embedding makes it possible to store both semantic and syntactic word information from an unannotated corpus and constructs a vector space in which word vectors are closer together if the words occur in the same linguistic contexts, i.e. if they are recognized as semantically more similar (according to the hypothesis of distributional semantics). 
In a stricter definition, word embedding is an overall term for a set of modeling techniques in natural language processing in which words or phrases of a vocabulary are mapped into vectors of real numbers. 
We have analyzed the tweets collected (as text) with the word embedding tools, to see, the most used words (which may suggest the main topics) and the most related words among the various groups.

\section{Results} 
\begin{figure} 
     \centering
     \includegraphics[width=1\linewidth]{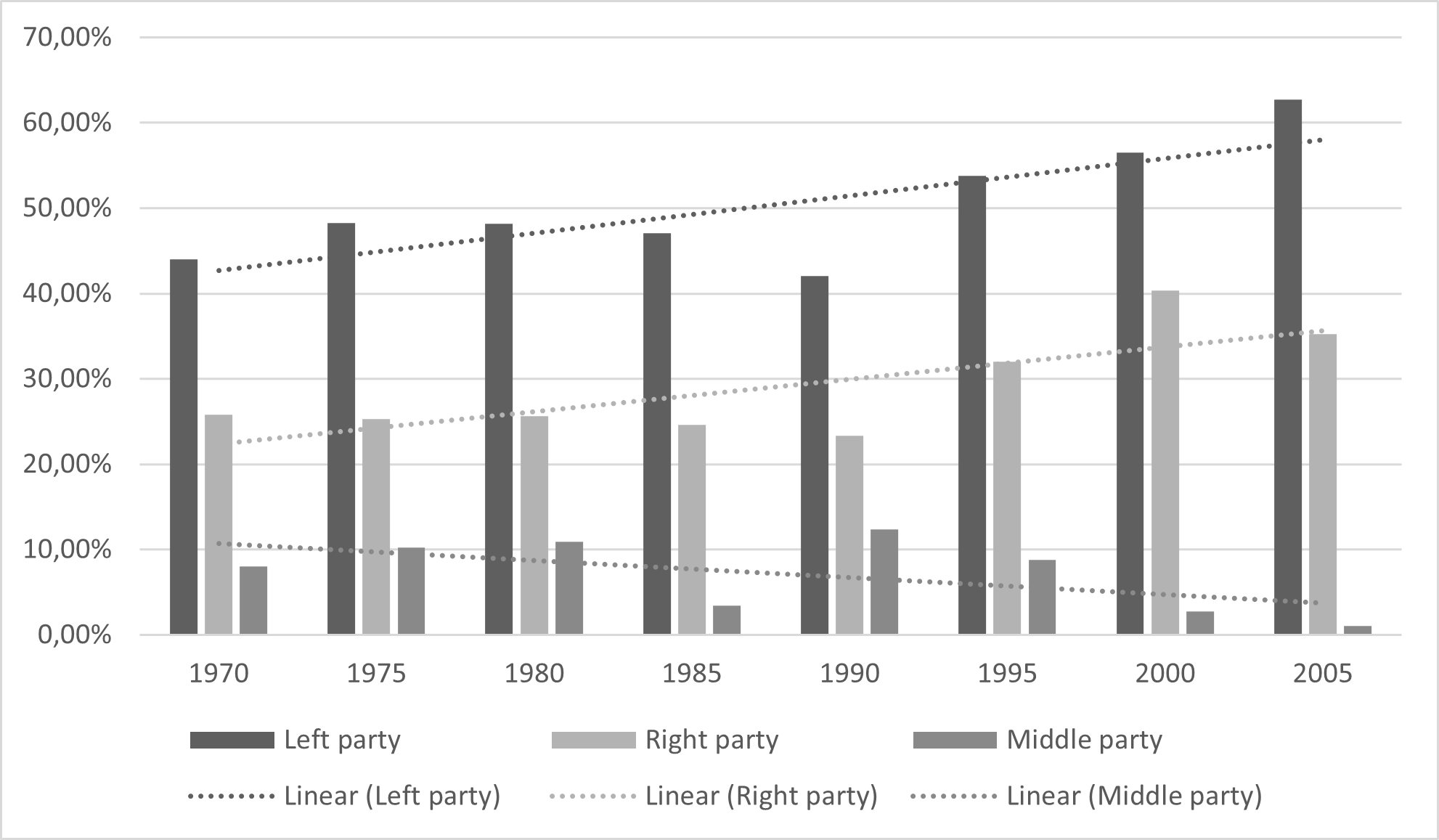}
     \caption{Election between 1970-2005}
\label{FIG:1}
\end{figure}

The Emilia Romagna regional election background, was, indeed a test for the national government and for the left-wing against right-wing national leadership. The Emilia Romagna - given the election results \cite{R1} and as shown in figure \ref{FIG:1}  - is a stronghold region for the left-wing political party in Italy, and for the first time, this region can be governed by the right political party, with the former leader Matteo Salvini (Leader of Lega Nord). 

   \begin{figure} 
     \centering
     \includegraphics[width=1\linewidth]{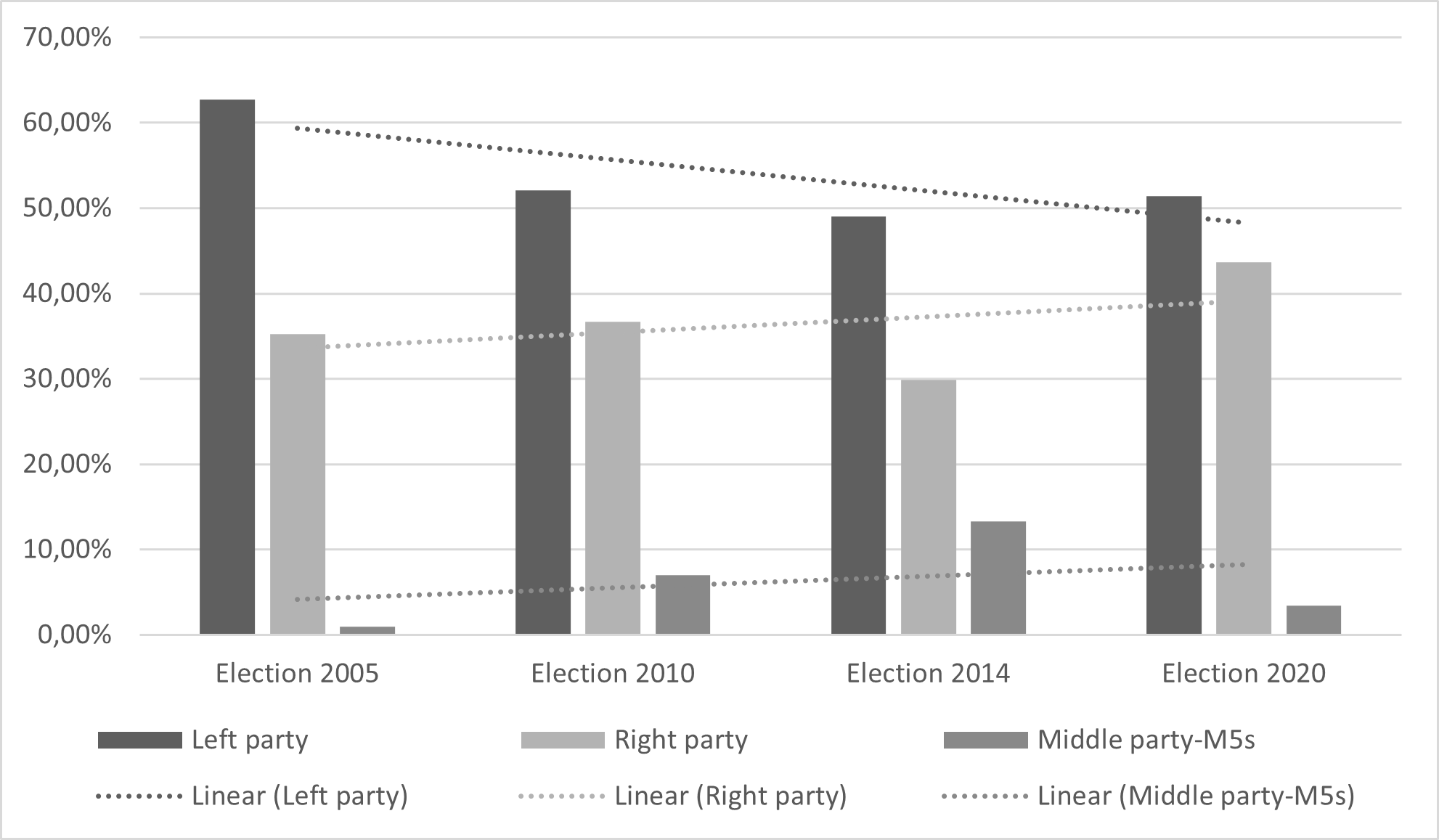}
     \caption{Election between 2005-2020}
\label{FIG:2}
\end{figure}

The event became a very important one, both by the amount of activity in social networks \cite{giuffrida2021analyzing}  \cite{russo2022entropy}, but also for the political and cultural left-wing political party \cite{R1}, because, as shown in figure \ref{FIG:2}, since the election of 2005 the right-wing party has continuously reduced the distance between both. So, if the left party lost, the election would certificate not only the leadership of Matteo Salvini as the leader with the most agreement throughout Italy, but also in the regional institutions of Emilia Romagna (which have always dealt with left-wing culture) and also in national institutions. 

The data analysis shows that 50\% of the collected posts refer to the new local movement (known as 6000sardine), but the network analysis highlights that the major and important hub is instead the leader of the populism and Euro-skepticism party. This contrast shows the reaction of social dynamics to populism propaganda during the election. The 6000sardine movement has preferred to react to opposing political propaganda (given the various populist events) than to voicing the presence of the stronghold culture in the Emilia-Romagna area. This means that the 6000sardine wasn't a stronghold culture offshoot, born to protect themselves, but a social reaction against the populism and Euro-skepticism presence in a digital environment. 

\begin{figure} [b]
     \centering
     \includegraphics[width=1\linewidth]{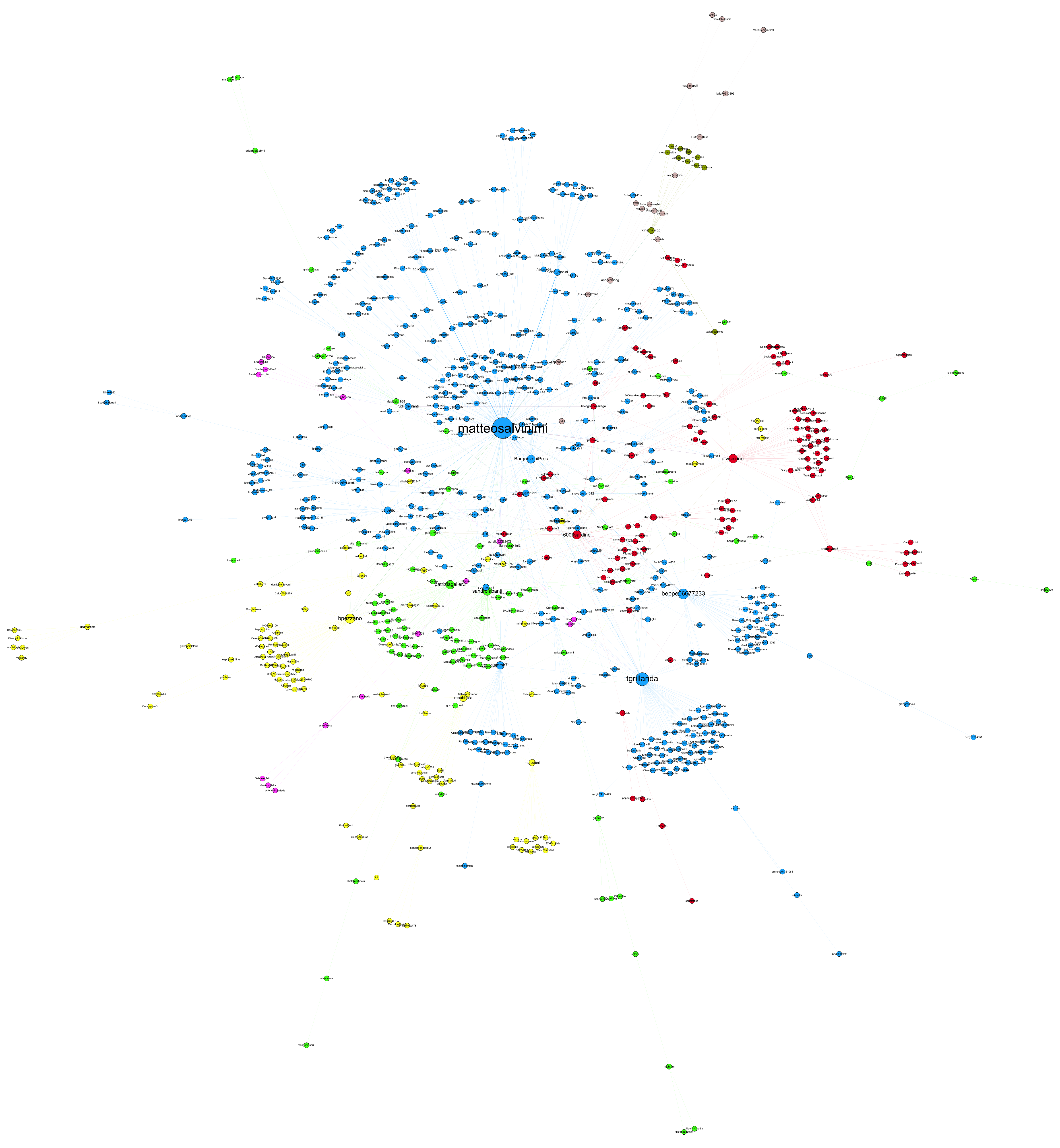}
     \caption{Populist and Sardine Network}
\label{FIG:3}
\end{figure}

\subsection{Network results}

The first result of this paper is related to the networks themselves, shown in Figure \ref{FIG:3}. The statistics of all networks are shown in Table \ref{Tab:2}.
The names of the nodes in the pictures are in proportion to each other. In this case, nodes with more links will have larger names given their importance. 

The network shows different characteristics: 1) For the colors for each community, we chose blue for the Populists, red for the Sardine Movement, yellow for M5s and journalists (the algorithm could not distinguish between them), while the other colors are various scattered communities; 2) Networks have 14 different communities (divided into populist, neutral, and pro), and 10 different hubs; \\
Of the 10 hubs (only selected if a minimum number of 30 likes were present and connected to the others through citations and replies), 6 of them belonged to the populist party, and 4 were against it.

\begin{figure} [htb]
    \centering
    \includegraphics[width=1\linewidth]{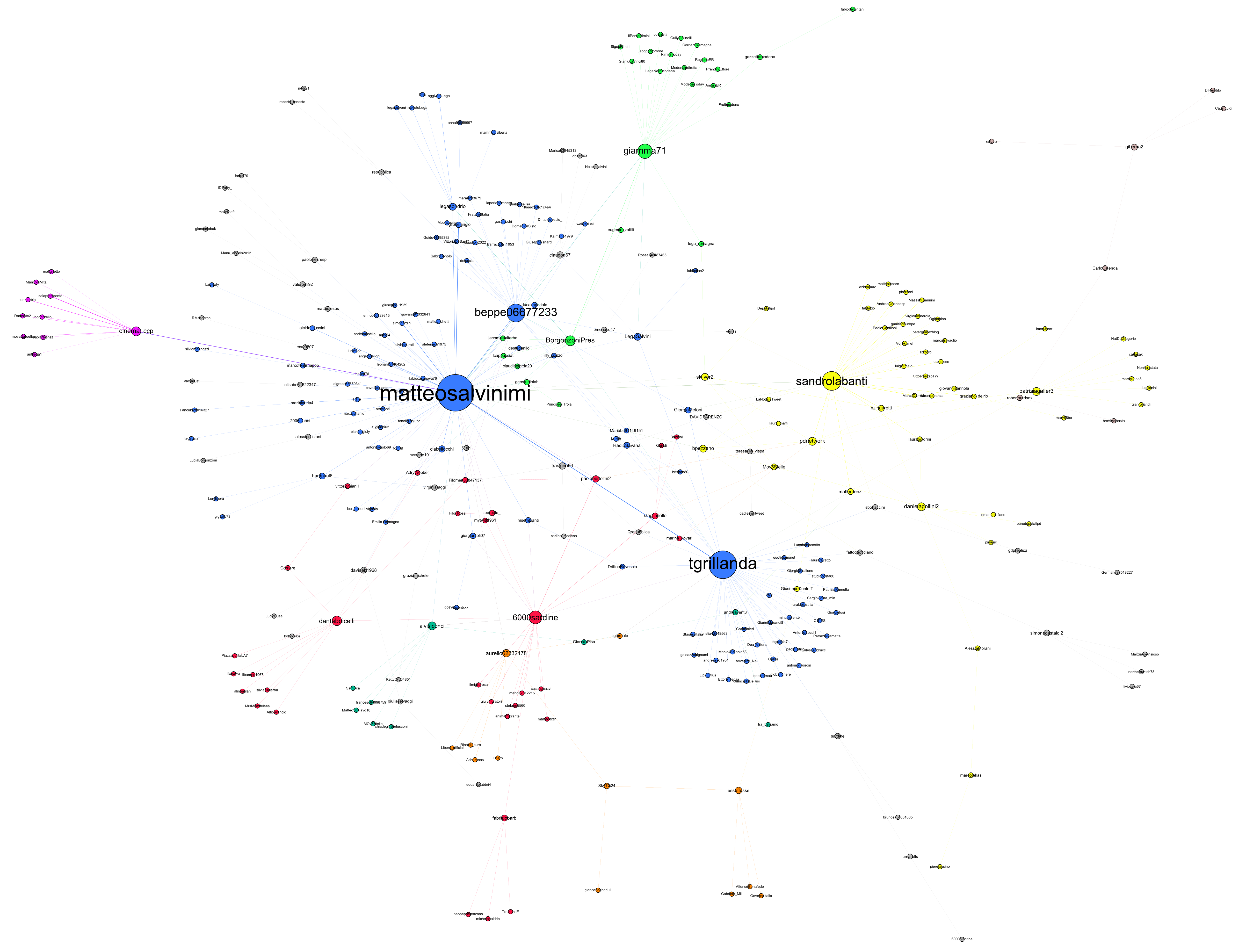}
    \caption{Populist party}
    \label{FIG:4}
\end{figure}

\begin{figure} [htb]
    \centering
    \includegraphics[width=1\linewidth]{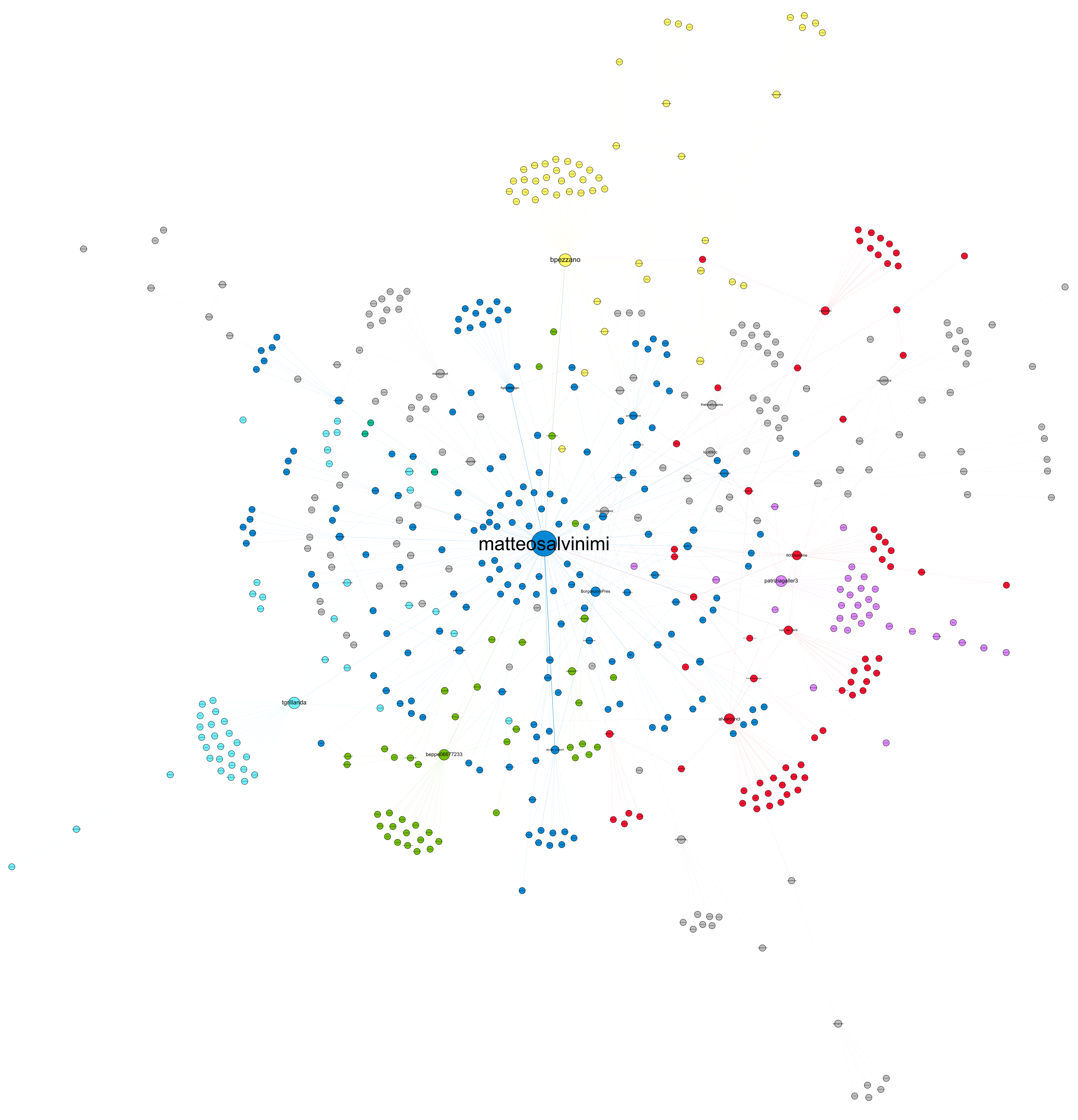}
    \caption{Sardine movement}
    \label{FIG:5}
\end{figure} 


\begin{table}  [t]
\centering
    \begin{tabular}{ |l| c|c|c|}
    \hline
    Statistic  & Both & Populist & Sardine \\
    \hline
    N. of nodes & 926 & 403 & 647\\
    N. of links  & 1104 & 422 & 749\\
    N. of community & 34 & 12 & 20 \\
    Modularity & 0.533 &  0.636 & 0.58 \\
    Density & 0.0026 & 0.05 & 0.035 \\
    Mean degree & 0.015 & 0.025 & 0.02 \\
    Mean betweenness & 0.012 & 0.024 & 0.4 \\
    \hline
    \end{tabular}
    \vspace{5pt}
\caption{Networks statistics}
\label{Tab:2}
\end{table}

The main hub of the populist network is the leader of the populist party (Matteo Salvini), as shown in Figure \ref{FIG:4}.


The same network also shows a few hubs close to the leader. This behaviour is uncommon during the election because usually most of the attention should be focused on the candidate (Lucia Borgonzoni), but she has a marginal role. Most of the attention was focused on the leader of the populist party. We may find an explanation for that in the leader's various populist actions during the election campaign, where he walked into an alleged drug dealer's house asking if he was dealing drugs \cite{spaccia}, or when he exploited a criminal case involving minors \cite{Bibbiano}. 

For these reasons, the Sardine Movement emerged, namely to counter the populist phenomenon. In fact, as Figure \ref{FIG:5} shows, the main hub of the network is exactly the leader of the Populist group, Matteo Salvini. This can be explained by what was said earlier, namely, the Sardine Movement was born as an \textit{Anti-Populist movement}, and not as an aid to leftist parties. 
In fact, the network characteristics show very few hubs, as they are all focused on the populist leader (the main one). Even in this case, there is no focus on the candidate (leftist) during the election.


These network dynamics mark that the populist party has fought the stronghold culture and that at the same time has also raised an attack against the 6000sardine movement. Both networks show a Barabasi Complex network  \cite{albert2002statistical} for populist and sardine networks.

\subsection{Word-embedding Result}
The second result is related to word embedding networks. The right-wing leader party Salvini was the principal hub in both networks; this evidences (again) that Salvini's role-action was the main topic during the election campaign, and not Emilia Romagna itself, with their respective political parties, leaders, and political programs/projects. The word embedding score and main topic/words used, both for Sardine and populist, is shown in Table \ref{tab:5}. 

\begin{table} [ht]
    \centering
    \begin{tabular}{ |l|| c|c|c|}
    \hline
    Words  & Sardine movement & Populist party & all\\
    \hline
    Salvini &  15 & 2 & 19\\
    Citofono (intercom) & 10 & -  & - \\
    Bambola-gonfiabile (sex-doll) & 10 & - & - \\
    Piazza & 5 & -  & 5\\
    Sardina & - & - & 5\\
    Stato &- & 3 & 5 \\
    PD & & 4 & - \\
    Italia & - & - & 4 \\
    Sardine  &- & 2 & -\\
    Governo & - & 2 & 3\\
    Gente& - &2 & -\\
    Lavoro & - & - & 3 \\
    Emilia & - & - & 3 \\
    Sinistra & - & - & 3\\
    \hline
    \end{tabular}
    \vspace{5pt}
\caption{Word-embedding Sardine movement}
\label{tab:5}
\end{table}




\section{Conclusion}
The present paper draws attention to how computational social science and network science can explain both the complex dynamics of controversial and challenging topics. The digital ecosystem does not only evolve communication on multinetwork (like TV, social media, and newspapers) but also provides the social researcher with useful data to explain social-complex dynamics.


In our work, we have shown how by analyzing political groups during important events, it is possible to show some characteristics of the social and political dynamics of various groups. For example, during the political election in Emilia Romagna, a large presence was shown both in social and also in more classical media, given the various behaviours of the leader Matteo Salvini, which triggered social mobility. 
Our work was mainly based on the analysis of networks, and words used during the election campaign, showing how a careful analysis of groups through networks, can give important information about the current event. 
In fact, none of the candidates was ever a main hub in the networks, while much more attention was paid (in both networks) to the actions of the populist leader. 

\section{Declarations}
The authors declare no conflict of interest.\\
The authors disclosed receipt of the following financial support for the research, authorship, and/ or publication of this article: This project has not received funding from the University of Catania.

\section{Authors contribution}
All Authors have contributed equally. 

\bibliography{biblio.bib}

\end{document}